\documentclass[%
 reprint,
 amsmath,amssymb,
 superscriptaddress,
 aps,pra
]{revtex4-2}

\usepackage[table]{xcolor}

\usepackage{siunitx}
\usepackage{graphicx}
\usepackage{dcolumn}
\usepackage{svg}
\usepackage{bm}
\usepackage[unicode=true, colorlinks=true, citecolor={blue!80!black}, urlcolor={blue!50!black}, linkcolor = {blue!80!black}]{hyperref}
\usepackage{easyReview}
\usepackage{adjustbox}
\usepackage{xcolor}
\usepackage[T1]{fontenc}
\usepackage{blindtext}
\usepackage{braket}
\usepackage{gensymb}
\usepackage{booktabs}
\usepackage[euler]{textgreek}

\begin{document}

\preprint{QuTech/AndersenLab}

\title{Coherence limitations of a Fourier-engineered cos(2\texorpdfstring{\textit{\textphi}}{phi}) transmon qubit}

\author{Nataliia K. Zhurbina}
\email{nkzhurbina@gmail.com}
\author{Siddharth Singh}
\author{Lukas J. Splitthoff}
\author{Eugene~Y.~Huang}
\author{Figen Yilmaz}
\author{A. Mert Bozkurt}
\author{Christian Kraglund Andersen}
\email{c.k.andersen@tudelft.nl}

\affiliation{QuTech and Kavli Institute of Nanoscience, Delft University of Technology, Delft 2628 CJ, The Netherlands}

\date{\today}

\begin{abstract}

Intrinsically protected superconducting qubits are a promising route toward enhancing coherence times and advancing hardware towards applications in quantum computing. The $\cos(2\varphi)$ qubit achieves protection against qubit relaxation by allowing only the coherent tunneling of pairs of Cooper pairs, resulting in Cooper-pair parity symmetry and thereby suppressing charge-induced errors. 
In this work, we experimentally realize a $\cos(2\varphi)$ qubit by Fourier engineering the energy-phase relation in a multi-junction superconducting circuit. Using an interference-based architecture, we are able to suppress the odd harmonics of an effective qubit potential and we observe good agreement between the measured transition spectrum and the effective theoretical model. 
We further investigate the energy relaxation time as a function of external flux and find that the qubit lifetime at the flux symmetry point is limited by $1/f$ flux noise. This strong sensitivity arises from residual fluctuations in the first harmonic, which possesses a large prefactor despite being nominally canceled. In contrast, a fluxonium qubit with a similar energy spectrum and noise amplitude is less affected by flux noise, highlighting a key challenge for interference-based protection schemes.

\end{abstract}

\maketitle

\section{Introduction}

The future realization of quantum computing with superconducting circuits will rely on active quantum error correction (QEC) to mitigate decoherence and operational errors by encoding logical states into many physical qubits. However, the overhead for implementing QEC requires hundreds of physical qubits per logical qubit~\cite{google2025quantum, Google2023, Fowler_2012, IBMHardware}. A complementary approach is to design qubits with intrinsic noise protection, which would significantly reduce the overhead for QEC~\cite{Gyenis_2021}. Protected qubits generally achieve suppressed error rates by encoding quantum information in systems with Hamiltonians possessing symmetries that restrict the allowed decoherence channels. In such systems, logical states are encoded in distinct symmetry sectors, often characterized by conserved quantities such as parity. As a result, noise-induced transitions between these states are strongly suppressed since they would require symmetry-breaking perturbations.

The most widely used superconducting qubit is the transmon qubit~\cite{Transmon}. In a conventional transmon qubit, a single Josephson junction is shunted with a large capacitor. The Josephson junction is described by a $\cos(\varphi)$ energy potential, where $\varphi$ is the superconducting phase difference across the junction, representing the tunneling of single Cooper pairs. The transmon qubit is protected against dephasing from charge noise because its Josephson energy is significantly larger than its charging energy. However, the charge dipole moment between qubit states remain large and, thus, the transmon is not protected against energy decay. In contrast to the Josephson junction, an element with a $\cos(2\varphi)$ potential would correspond to the coherent tunneling of \emph{pairs} of Cooper pairs. A Hamiltonian with a $\cos(2\varphi)$ potential therefore conserves Cooper-pair parity and couples only charge states within the same parity sector. As such, a qubit can have states encoded within opposite parity sectors in order to decouple it from the superconducting charge operator. In other words, we expect this qubit to be protected against energy decay. Several approaches have been proposed to realize such a $\cos (2\varphi)$ potential, including protected rhombus chains \cite{Protected_Rhombus_Bell}, bi-fluxon tunneling in fluxonium circuits \cite{Bi-fluxed_Fluxonium_Wael}, the $0-\pi$ qubit \cite{Zero-pi_Brooks_2013, Catelani2012_QP_noise, Gyenis_2021_0_pi}, and the kinetic interference co-tunneling element (KITE) \cite{Smith2020, roverch2026experimentalcos2phi, nguyen2025gridstatesqubit}.

An effective $\cos(2\varphi)$ potential can also be realized through a superconducting quantum interference device (SQUID) loop biased at half a magnetic flux quantum ($\Phi_0/2$), where odd harmonics cancel while even harmonics remain. Of course, this requires the presence of higher harmonics in the junction potential and, thus, interference-based implementations have been mostly demonstrated using various superconductor -- normal metal -- superconductor (SNS) junctions, such as InAs nanowires \cite{Larsen2020, feldsteinbofill2026}, planar InAs systems \cite{InAs_planar_cos2phi_Zhang}, Ge/SiGe two-dimensional electron gases \cite{Ge-based_cos2phi_Leblac}, and graphene-based junctions \cite{Graphene_CPR_Nanda_2017, Graphene_cos2phi_Messelot}.
In this work, in contrast to using SNS junctions, we implement higher-mode engineering with a series of Josephson junctions, enabling Fourier control over the energy–phase relation~\cite{Bozkurt_2023, Bozkurt_2023_valla, shagalov2025}. Our design incorporates flux tunability to precisely control the effective higher harmonics of the circuit, allowing us to tune between a transmon-like potential and the protected $\cos(2\varphi)$ regime. The primary experimental objective is to tune the device between these regimes and validate the accuracy of an effective single-mode Hamiltonian.

\clearpage  

\begin{figure}[t]
    \centering
    \includegraphics[width=0.99\linewidth]{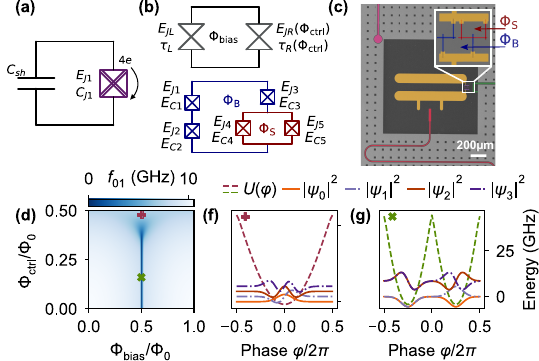}
    \caption{{Qubit circuit and device design}.
    (a) Idealized circuit implementing a $\cos(2\varphi)$-protected qubit. The cross element enables the tunneling of pairs of Cooper pairs, generating an effective $\cos(2\varphi)$ potential. The tunneling element is additionally shunted by a large capacitor.
    (b) Top: Interference-based implementation of the $\cos(2\varphi)$ element using SNS junctions. Bottom: Schematic of the fabricated device, where the left SQUID arm forms an effectively fixed-energy SNS junction using two Josephson junctions in series. The right arm also implements an effective SNS junction with a single Josephson junction in series with a SQUID loop, such that the effective SNS junction parameters are tunable.
    (c) SEM image of the tunable qubit. The charge line is shown in pink, the readout resonator in red, and the flux bias line in green. The shunt capacitor is highlighted in yellow, and the junction structure in red and blue.
    (d) Qubit transition frequency $f_{01}$ as a function of the fluxes threading the two SQUID loops independently. Selected points indicate the qubit operating in the transmon (red) and $\cos(2\varphi)$ (green) regimes.
    Panels (f) and (g) show the corresponding potentials and wavefunctions for these operating points. The wavefunctions are shifted upwards by the eigenenergy of the corresponding state.}
    \label{fig:device}
\end{figure}

\section{Results}

\subsection{Device design}

We aim to design a transmon qubit with an element that produces a $\cos(2\varphi)$ potential, see Fig.~\ref{fig:device}(a). As mentioned, one approach is to use a SQUID with SNS junctions, shown in Fig.~\ref{fig:device}(b, top). In the limit of a single short transmission channel, each SNS junction has an energy-phase relation (EPR) of the form $\Delta\sqrt{1 - \tau \sin^2(\varphi/2)}$ ~\cite{pitavidal2025}, where $\Delta$ is the effective superconducting gap and $\tau$ is the transparency of the transmission channel. Since the EPR is $2\pi$-periodic, it can be decomposed as a Fourier series in the form
\begin{equation}
  \Delta\sqrt{1 - \tau \sin^2(\varphi/2)} = \sum_n E^{n}_J \cos(n\varphi), \label{eq:sns_potential}
\end{equation}
where $E_J^n$ are the energies associated with each harmonic. The amplitude of the harmonics depends on the transparency $\tau$ and decreases for higher-order terms. It has also been shown that a conventional superconductor-insulator-superconductor (SIS) Josephson junction exhibits higher harmonics, however, their amplitudes are typically much smaller \cite{kim2025emergentharmonics, Willsch_2024}. For a transmon-like circuit with a SQUID loop of two SNS-junctions biased at half a magnetic flux quantum ($\Phi_{\rm bias} = \Phi_0/2$), the Hamiltonian takes the form
\begin{align} 
        H ={}&\, 4E_C(\hat{n} - n_g)^2  - \!\!\sum_{n = 2k+1} [E^{ n}_{JL} -E^{n}_{JR}]\cos(n\hat{\varphi}) \nonumber \\
         &- \sum_{m= 2k} [E^{n}_{JL} + E^{n}_{JR}]\cos(m\hat{\varphi}),\label{eq:SNS_Squid_at_0.5}
\end{align}
where $E_C$ is the charging energy of the transmon and $E_{JL}^n$, $E_{JR}^n$ are the Fourier coefficients of the left and right junctions, respectively. By tuning the energy of one of the junctions such that the SQUID becomes symmetric, the resulting Hamiltonian contains only even harmonics, with the dominant contribution arising from the $\cos(2\varphi)$ term.
An alternative approach to implement the potential in Eq.~\eqref{eq:sns_potential} is to use a series of Josephson junctions to Fourier engineer the desired potential~\cite{Bozkurt_2023, Bozkurt_2023_valla}. It turns out that an SNS energy-phase relation can be recreated using two SIS junctions in series, under the assumption that we can neglect the charging energy of the island between them \cite{Bozkurt_2023}. Such a design can be seen as a minimal rhombus qubit \cite{Protected_Rhombus_Bell} with additional tunability.
Taking two junctions in series with Josephson energies $E_{J1}$ and $E_{J2}$ and phase drops $\varphi_1$ and $\varphi_2$ across each junction, we find an effective energy–phase relation 
\begin{equation}\label{eq:effective_EPR_bowtie}
    U_{\bowtie}(\hat{\varphi})= - E_{J\Sigma}\sqrt{1 - \tau \sin^2(\hat{\varphi}/2)},
\end{equation}
where the effective energy scale and transparency are
 \begin{equation}
    E_{J\Sigma} = E_{J1} + E_{J2}, ~\tau = \frac{4E_{J1}E_{J2}}{(E_{J1} + E_{J2})^2},
\end{equation}
and the total phase drop is $\varphi = \varphi_1 + \varphi_2$. 

An additional contribution to this simplified model arises from consideration of the internal modes of the multi-junction structure, as discussed in Ref.~\cite{shagalov2025, jakobsen2026_double_junctions}. Proper accounting of the internal mode energy is necessary for an accurate description of the system. Following the Born-Oppenheimer approximation of Ref.~\cite{shagalov2025}, the internal mode energy is given by
\begin{equation} \label{eq:internal_mode}
U_{\bowtie}^\mathrm{int}(\varphi)  ={E_{J\Sigma}}\sqrt{\frac{2{E_C^\mathrm{int}}}{{E_{J\Sigma}}}\sqrt{1 - \tau \sin^2(\hat{\varphi}/2)}}, 
\end{equation}
where $ E_C^\mathrm{int} = {e^2}/[2(C_{J1} + C_{J2})]$ is the charging energy of the internal mode.

In our device, we use a SQUID loop with a fixed left arm with two Josephson junctions in series and a tunable right arm with a single junction in series with a smaller SQUID loop, see the circuit schematic in Fig.\ref{fig:device}(b, bottom) and device image in Fig.~\ref{fig:device}(c). The small loop enables the tuning of the effective transparency of the right arm, allowing a continuous transition from a conventional transmon to a regime dominated by a $\cos(2\varphi)$ potential.
We embed this SQUID loop between two large capacitor pads, designed to yield a charging energy of $E_C/h = \qty{0.21}{\GHz}$. To mitigate charge-dispersion in the $\cos(2\varphi)$ regime, the junction energies are chosen to be comparatively large, with $E_J^{n=2}/E_C \gg 50$. The capacitor pads are also capacitively coupled to a readout resonator and a microwave drive line. The flux biases are controlled with a local flux-bias line (FBL) and a global magnetic flux provided by an off-chip coil. The capacitive structures, as well as the coplanar waveguide lines, are defined in a Nb/Ta bilayer. The Josephson junctions are fabricated in a Manhattan-style geometry using an aluminum lift-off process, additional fabrication details are provided in (Sec.~\ref{sup:Fabrication}). 

To model the system, we describe our qubit by the effective Hamiltonian
\begin{equation} \label{eq:Hamiltonian}
    \begin{split}
        H ={}& 4E_C(\hat{n} - n_g)^2 -  E_{JL}\sqrt{1 - \tau_{L}\sin^2\dfrac{\hat{\varphi}}{2}} \\
        &+ E_{JL}\sqrt{\frac{2{E_{CL}^\mathrm{int}}}{{E_{JL}}}\sqrt{1 - \tau_{L}\sin^2\dfrac{\hat{\varphi}}{2}}}\\
        &-  E_{JR}\sqrt{1 - \tau_{R}\sin^2\dfrac{\hat{\varphi}-\varphi_{\rm B}-\delta}{2}}\\
         &+ E_{JR}\sqrt{\frac{2{E_{CR}^\mathrm{int}}}{{E_{JR}}}\sqrt{1 - \tau_{R}\sin^2\dfrac{\hat{\varphi}-\varphi_{\rm B}-\delta}{2}}}
         \,,
    \end{split}
\end{equation}
where $\varphi_{\rm B} = 2\pi\Phi_{\rm B}/\Phi_0$ is the flux bias of the ``big'' SQUID loop (see Fig.\ref{fig:device}(b, bottom)) and $E_{CL/R}^\mathrm{int}$ is the charging energy of the internal mode on the left/right arm. An additional phase offset $\delta = \varphi_{\rm S}/2 + \arctan(d \tan(\varphi_{\rm S}/2))$ is introduced by the ``small'' SQUID loop in the right arm, where $\varphi_{\rm S} = 2\pi\Phi_{\rm S}/\Phi_0$ denotes its external flux bias.

The effective Josephson energies of the left and right arms are given by $E_{JL} = E_{J1} + E_{J2}$ and $E_{JR} = E_{J3} + E_{J45}^\mathrm{eff}$, respectively, where
\begin{equation}
E_{J45}^\mathrm{eff} = E_{J45}\sqrt{\cos(\pi\varphi_{\rm S})^2 + d^2\sin(\pi\varphi_{\rm S})^2}
\end{equation}
with  $E_{J45} =E_{J4} + E_{J5}$ and $ d = (E_{J4} - E_{J5})/(E_{J4} + E_{J5})$, derived from the small SQUID loop \cite{Transmon}. Due to the presence of the phase offset, the sweet spot in $\Phi_{\rm B}$ is shifted from half a flux quantum proportionally in $\delta$. For the rest of this manuscript, we use instead a redefined flux coordinate $\Phi_{\rm bias} = \Phi_{\rm B} - \delta(\Phi_0/2\pi)$. In this way, the symmetry point is fixed at $\Phi_{\rm bias}/\Phi_0 = 0.5$, while the control flux is identified with the small-loop flux, $\Phi_{\rm ctrl}~=~\Phi_{\rm S}$, corresponding to the interference-based SNS device in Fig.~\ref{fig:device}(b, top). Accordingly, the Hamiltonian used in the subsequent analysis corresponds to Eq.~\ref{eq:Hamiltonian}, taking the $\delta$ offset into account only implicitly.

Reducing the multi-junction Hamiltonian to an effective SNS interference qubit introduces a single effective phase variable for each composite arm, rather than assigning independent phases to each junction. As a result, the model is significantly simplified while retaining the relevant higher harmonics of the qubit structure. Specifically, this model captures the asymmetry between the two arms of the big SQUID loop as being controlled by $\Phi_{\rm ctrl}$. When $\Phi_{\rm ctrl}$ is biased close to $\Phi_0/2$, the right arm is effectively suppressed, resulting in a transmon-like regime as shown in Fig.~\ref{fig:device}(e). Conversely, when the system is tuned toward a symmetric configuration where both arms contribute equally, the potential develops a double-well structure corresponding to the protected regime, see Fig.~\ref{fig:device}(f).

\subsection{Spectroscopy}

\begin{figure}[t]
    \centering
    \includegraphics[width=0.99\linewidth]{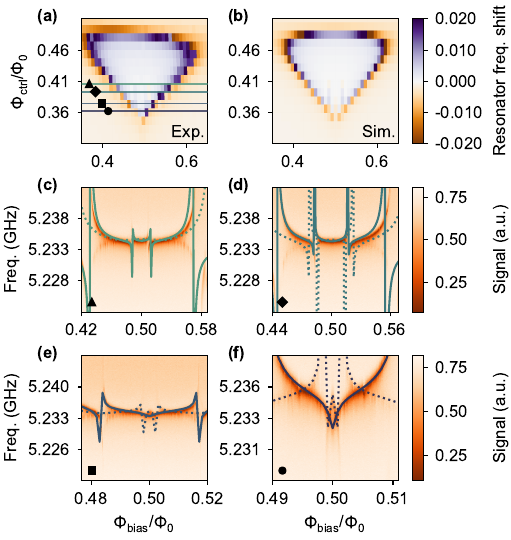}
    \caption{
    (a) Measured resonator frequency shift as a function of the external fluxes, after cross-talk calibration. Horizontal traces (with markers) correspond to additionally measured line cuts, ordered from top to bottom with increasing $\cos (2\varphi)$ contribution.
    (b) Simulated resonator frequency shift using Eq.~\eqref{eq:res_shift}.
    (c–f) Line cuts of (a) for $\Phi_{\rm ctrl}$ equal to 0.406\,$\Phi_{0}$, 0.395\,$\Phi_{0}$, 0.378\,$\Phi_{0}$, and 0.367\,$\Phi_{0}$, respectively, overlaid with simulated resonator shifts. Solid lines indicate transitions from the ground state, while dotted lines correspond to transitions from the first excited state.
    }
    \label{fig:Resonator_spec}
\end{figure}

The coupling between the resonator and the qubit (tuned by $\Phi_{\rm bias}$ and $\Phi_{\rm ctrl}$) introduces a flux-dependent resonator frequency shift. To measure this shift, we perform spectroscopy by applying a single microwave tone to a transmission line capacitively coupled to the readout resonator. By sweeping the probe frequency through the transmission line around the typical resonator frequency, we measure the transmitted signal and, from a Lorentzian fit, we extract the resonator frequency as a function of $\Phi_{\rm bias}$ and $\Phi_{\rm ctrl}$, see Fig.~\ref{fig:Resonator_spec}. During these measurements, we compensate for the linear cross-talk between an external coil and the on-chip flux-bias line to independently tune $\Phi_{\rm bias}$ and $\Phi_{\rm ctrl}$, see Supplementary Information (Sec.~\ref{Sup: Cross-talk calibration}).

To further understand the resonator response, we compute the resonator frequency shift using the expression
\begin{equation} \label{eq:res_shift}
    \begin{split}
  \delta\omega_\mathrm{res} = g^2\sum_{i\neq0} |\bra{i}\hat{n}\ket{0}|^2\frac{-2\omega_{i0}}{\omega_{i0}^2-\omega_\mathrm{res}^2}
    \end{split}    
\end{equation}
where $g$ is the geometric coupling between the qubit and the resonator. The summation over $i$ accounts for all excited states of the qubit, and $\omega_{i0}$ refers to the angular frequency of the qubit ground state subtracted by the frequency of state $i$. Here, $\omega_\mathrm{res}$ is the bare resonator angular frequency. 
The frequency shift calculated from Eq.~\eqref{eq:res_shift} shows excellent agreement with the experimental data. The matrix element and eigenfrequencies in Eq.~\eqref{eq:res_shift} were obtained from the numerical solution of Eq.~\eqref{eq:Hamiltonian}. The system parameters are extracted from two-tone spectroscopy measurements, as discussed later.

To highlight signatures of the protected regime associated with the expected double-well potential, we show the resonator frequency shift for four flux-bias values in Fig.~\ref{fig:Resonator_spec}(c–f). In these panels, the data are shifted such that the symmetry point is centered at $\Phi_{\rm bias}/\Phi_0 = 0.5$, i.e., by compensating for the phase offset $\delta$. The avoided crossing between the resonator and the $\ket{0}\leftrightarrow\ket{1}$ transition shifts toward the symmetry point while its magnitude decreases. In a double-well potential, this transition corresponds to an intra-well excitation that is suppressed near the symmetry point. Along with the frequency shift associated with the qubit ground state $\ket{0}$, we observe an additional resonance in the spectroscopy data, in particular seen in Fig.~\ref{fig:Resonator_spec}(f). This additional feature agrees well with the expected resonator frequency shift for the qubit in the $\ket{1}$ state, as calculated from Eq.~\eqref{eq:res_shift} with $\ket{0}$ replaced by $\ket{1}$. It is most prominent in Fig.~\ref{fig:Resonator_spec}(f) near the symmetry point, where the qubit frequency is low, and the thermal population of the excited state becomes significant. The close agreement between the experimental data and the model indicates that the resonator–qubit interaction is accurately captured and that the model reliably reproduces the observed spectral features.

\begin{figure}[t]
    \centering
    \includegraphics[width=0.99\linewidth]{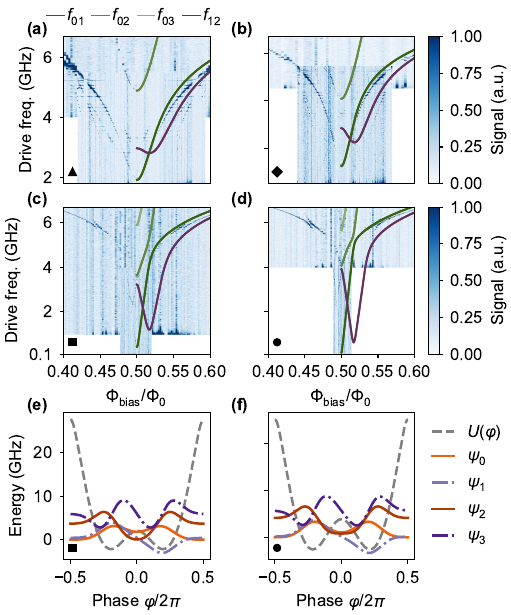}
    \caption{
    (a–d) Two-tone qubit spectroscopy corresponding to selected measurement points as a function of $\Phi_{\rm bias}$,
    see the annotations of 
    Fig.~\ref{fig:Resonator_spec}(a). Additionally, in solid lines, we show the fitted spectra. To avoid obscuring the data, the fit is only displayed for half of the flux axis.
    (e–f) Calculated potentials and wavefunctions based on the fits to the spectra in (c) and (d), respectively. 
    }
    \label{fig:Qubit_spec}
\end{figure}
 
Next, we perform two-tone spectroscopy measurements by applying a microwave tone to the microwave drive line and sweeping its frequency while driving a readout tone at the resonator frequency, see Fig.~\ref{fig:Qubit_spec}(a–d). 
In Fig.~\ref{fig:Qubit_spec}(a), the $f_{01}$ transition follows a U-shaped profile characteristic of an asymmetric-transmon-like potential with a single minimum. As $\Phi_{\rm ctrl}$ is decreased and the SQUID approaches a symmetric configuration, the spectrum evolves into a V-shaped profile, as expected when the first harmonic is being suppressed, as was also observed in previous works \cite{kim2025emergentharmonics, feldsteinbofill2026, Larsen2020}. The resulting spectrum was fitted by numerically diagonalizing the Hamiltonian in Eq.~\eqref{eq:Hamiltonian}. The charging energy $E_C = \qty{0.21}{\GHz}$ is fixed based on electrostatic simulations, and we assume it does not differ significantly to the fabricated device so that only the junction energies remain as free parameters. For the fixed arm of the loop, the fitted parameters are $E_{J1}/h = \qty{42.49}{\GHz}$, $E_{J2}/h = \qty{53.9}{\GHz}$, while for the tunable arm they are $E_{J3}/h = \qty{88.11}{\GHz}$ and $E_{J4}/h = E_{J5}/h = \qty{35.73}{\GHz}$. The simplified circuit model accurately reproduces the transition spectrum, as evidenced by the fit in Fig.~\ref{fig:Qubit_spec}(a\nobreakdash-d).

Using the fitted parameters, we visualize the potential and corresponding wave functions at $\Phi_{\rm ctrl}/\Phi_0 = 0.378$ and $\Phi_{\rm ctrl}/\Phi_0 = 0.367$, see Figs.~\ref{fig:Qubit_spec}(e) and (f), respectively. We observe the presence of a double-well potential with the eigenstates existing as even and odd superpositions of wavefunctions localized in each well.
\begin{figure*}
    \centering
    \includegraphics[width=1\textwidth]{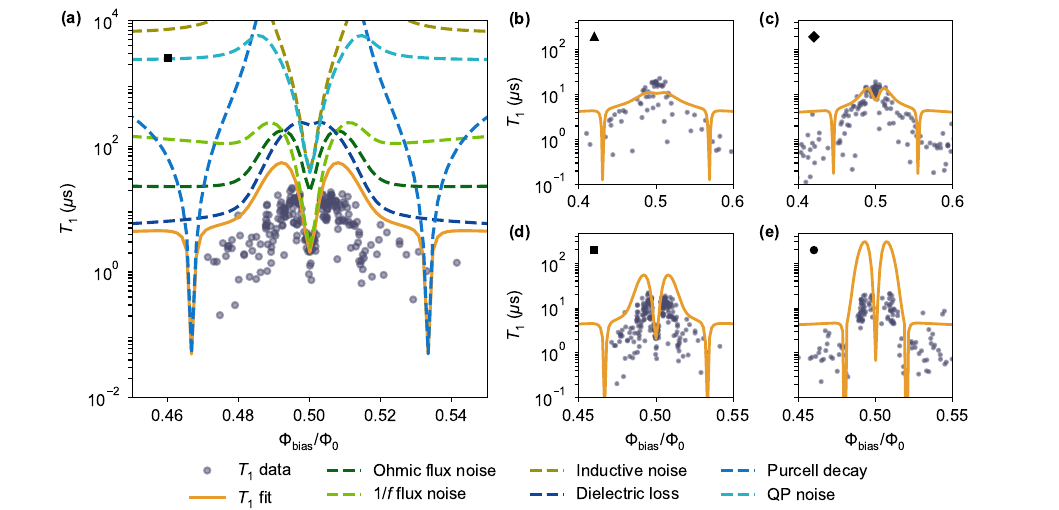}
    \caption{
    (a) Measured qubit lifetime, $T_1$, as a function of $\Phi_{\rm bias}$ for a fixed $\Phi_{\rm ctrl} = 0.378\,\Phi_0$. Dashed lines indicate the fitted contributions from individual noise sources (see legend), while the solid lines represent the total predicted $T_1$.
    (b-e) Lifetimes for four previously characterized bias points of $\Phi_{\rm ctrl}$, see 
    Fig.~\ref{fig:Resonator_spec}(a), overlaid with the total predicted noise contribution (solid).
    }
    \label{fig:T1}
\end{figure*}

\subsection{Characterization of qubit lifetime}

To investigate to what extent the qubit exhibits protection to energy relaxation, we measure the qubit lifetime as a function of the external flux bias $\Phi_{\rm bias}$. The energy relaxation measurements are carried out using standard time-domain techniques. The qubit is initialized in the $\ket{1}$ state with a $\pi$-pulse on the $\ket{0} \leftrightarrow \ket{1}$ transition, calibrated via Rabi oscillation measurements. We employ passive qubit reset, with a delay between measurements approximately five times longer than the $T_1$ time. In post-processing, we discard data points with large fitting uncertainties, see Supplementary Information (Sec.~\ref {sup:Measurement setup}). 

In Fig.~\ref{fig:T1}(a), we see that the qubit lifetime $ T_1$ increases as we approach the symmetry point. However, the lifetime decreases sharply at the flux bias point of $0.5\,\Phi_0$. To further quantify the flux dependence, we analyze the contributions from different noise sources. Noise can couple to the system through two conjugate quantum operators: phase $\hat{\varphi}$ and charge $\hat{n}$. Phase noise couples to the magnetic flux threading either SQUID loop and includes Ohmic flux noise and $1/f$ flux noise. Assuming that the internal mode associated with the middle island is highly lossy, we suggest that it gives rise to effective dissipation, which can be modeled as an inductive loss channel. Finally, charge noise arises from dielectric loss, quasiparticle (QP) poisoning, and Purcell decay due to coupling to the readout resonator.

\begin{table}[t]
\centering
\resizebox{\columnwidth}{!}{%
\begin{tabular}{@{}c|ccc@{}}
\toprule
\textbf{Noise source}   & \multicolumn{1}{c}{\textbf{Operator}} & \multicolumn{1}{c}{\textbf{Parameter}} & \textbf{Value}        \\ \midrule
Dielectric losses       & $2e\hat{n}$    & $Q_\mathrm{cap}$  & $1\times10^5$        \\
Purcell decay           & $\hat{n}$    & $g/2\pi$        & \qty{25}{\MHz}                 \\
Quasiparticle poisoning & $\sin(\hat{\varphi}/2)$ & $x_\mathrm{qp}$  & $7\times10^{-10}$  \\
Ohmic flux noise        & $\partial{\hat{H}}/\partial{\Phi}$  & $M_1, M_2$                        & $1800\,\Phi_0/A$         \\
$1/f$ flux noise        & $\partial{\hat{H}}/\partial{\Phi}$  & $A_{1/f}$                      & $1.5\times10^{-5}\,\Phi_0$ \\
Inductive noise         & $\frac{\Phi_0}{2\pi}\hat{\varphi}$ & $Q_\mathrm{ind}$                         & $5\times10^8$          \\ 
\bottomrule
\end{tabular}%
}
\caption{Noise sources considered in this work. Each noise source is characterized by an associated noise operator and a spectral density with a free fitting parameter. Details of the spectral density models are provided in the Supplementary Information (Sec.~\ref{sup: Noise}). The fitted parameter values are shown in the right column of the table.}
\label{tab:noise_params}
\end{table}

To quantify the corresponding relaxation rates, we model the system using numerical simulations of the effective Hamiltonian in Eq.~\eqref{eq:Hamiltonian} and evaluate the transition rates using Fermi’s golden rule. The decay rate from $\ket{1}$ to $\ket{0}$ is given by
\begin{equation}
    \begin{split}
\Gamma_{01} = \sum_\lambda \frac{1}{\hbar^2}|\bra{0} \hat{D}_\lambda\ket{1}|^2 S_\lambda(\omega_{10}),
    \end{split}    
\end{equation}
where $\hat{D}_\lambda$ is the noise operator associated with the noise source $\lambda$ and $S_\lambda(\omega_{10})$ is the noise spectral density evaluated at the qubit frequency. For example, dielectric loss couples to the charge operator and is parameterized by the capacitive quality factor $Q_\mathrm{cap}$. The corresponding operators and fitting parameters for the other noise sources are summarized in Table~\ref{tab:noise_params} and described in the Supplementary Information (Sec.~\ref{sup: Noise}).
In Fig.~\ref{fig:T1}(a), we show the contributions for each noise source. In Fig.~\ref{fig:T1}(b-e), the measured data are overlaid with simulations of the total noise contributions for four different values of $\Phi_{\rm ctrl}$. We generally observe good agreement between the predicted relaxation rates and the measured lifetimes.

Away from the symmetry point, the relaxation time is primarily limited by dielectric loss. However, we find that charge-based noise sources do not limit the measured lifetime at the symmetry point. For the $\cos(2\varphi)$ qubit, this realization is problematic since the qubit is designed to be protected against dielectric loss. Instead, the overall flux dependence of the qubit lifetime arises from a combination of Ohmic flux noise, $1/f$ flux noise, dielectric loss, and quasiparticle poisoning. In particular, at the symmetry point, we find that $1/f$ flux noise is the dominant noise source.

The strong influence of $1/f$ flux noise arises from the high sensitivity of the system to flux fluctuations that control the cancellation of the $\cos(\varphi)$ component. Fluctuations in flux restore the first harmonic, thereby reducing the effectiveness of its suppression. As mentioned in Ref.~\cite{messelot2026coherence}, in interference-based $\cos(2\varphi)$ implementations, the coherence time near the sweet spot can decrease for larger ratios $E_{J}/E_{C} \gg 100$, due to strong magnetic-field sensitivity, as is the case for our device. The detrimental impact of flux noise arises from transition rates that scale with the Josephson energy associated with the noise term, see discussion in Supplementary Information (Sec.~\ref{sup: Noise}). In our system, the relevant energy scale is set by the effective Josephson energy of either arm of the $\Phi_{\rm bias}$ SQUID loop, which remains on the order of \qty{40}{\GHz}. As discussed in Sec.~\ref{sec:fluxonium}, this large sensitivity to flux noise is a particular feature of our implementation of the $\cos(2\varphi)$ transmon. Therefore, despite the suppression of the first harmonic, the large overall energy scale of the circuit limits effective decoupling from flux noise.

Finally, we note that in Figs.~\ref{fig:T1}(d) and (e), the lifetime appears limited to around \qty{25}{\us} despite the noise models predicting higher values slightly away from the symmetry point. The origin of the observed ceiling in lifetime is currently unknown. To investigate this effect, we additionally account for transitions involving higher energy levels, which can be relevant closer to the symmetry point. This contribution reduces the expected coherence time but does not fully explain the observed saturation. A detailed analysis of the higher-level contributions, following the approach in Ref.~\cite{ateshian2025}, is provided in Supplementary Information (Sec.~\ref{Higher level contribution}).

\subsection{Comparison with a  fluxonium qubit}
\label{sec:fluxonium}

To highlight that the limited lifetime at the symmetry point is specific to our $\cos(2\varphi)$ implementation, we compare it with the more widely studied fluxonium qubit~\cite{Manucharyan_2009, Nguyen2019_Fluxonium, Feng22_Fluxonium, Somoroff23_Fluxonium, Nguyen22_Blueprint}.
The fluxonium circuit consists of a single Josephson junction with energy $E_J$ in parallel with a large superinductance characterized by energy $E_L$, and shunted by a capacitance with energy $E_C$. The Hamiltonian of the system is given by
\begin{equation} \label{eq:Fluxonium_ham}
H = 4E_C\hat{n}^2 - E_J\cos(\hat{\varphi} ) + \frac{1}{2}E_L(\hat{\varphi}- \varphi_\mathrm{ext})^2 
\end{equation}
where $ \varphi_\mathrm{ext} = 2\pi\Phi_\mathrm{ext}/\Phi_0$ is the external flux threading the loop formed by the junction and inductor. The fluxonium qubit is typically operated in the regime $E_L < E_J$ and $1 \lesssim E_J/E_C \lesssim 10$, such that the inductive energy introduces a parabolic potential modulated by the periodic $\cos (\varphi)$ term. In this case, when $\Phi_\mathrm{ext}/\Phi_0 = 0.5$, the potential is symmetric and the qubit frequency becomes first-order insensitive to dephasing via flux noise.

\begin{figure}
    \centering
    \includegraphics[width=0.5\textwidth]{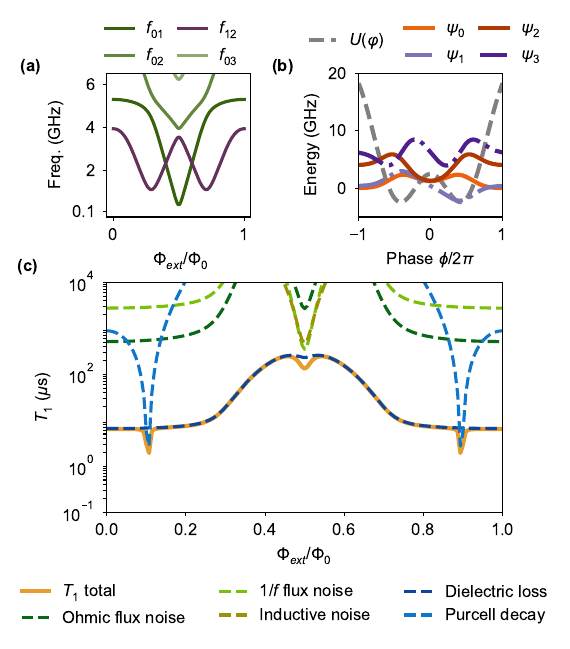}
    \caption{
    Fluxonium qubit spectrum and $T_1$ simulation.
    (a) Fluxonium qubit transition frequency spectrum.
    (b) Potential and wavefunctions of the fluxonium qubit at the symmetry point (half a flux quantum).
    (c) Simulated contributions of individual noise sources for the fluxonium qubit (dashed lines), using the same noise models and parameters extracted from fits to the measured $\cos(2\varphi)$ qubit, along with the total noise contribution (solid line).
    }
    \label{fig:Fluxonium}
\end{figure}

Here, we consider a fluxonium device with parameters $E_L/h = \qty{0.8}{\GHz}$  and $E_C/h = \qty{1}{\GHz}$, which are typical for a standard fluxonium circuit \cite{Nguyen2019_Fluxonium, Manucharyan_2009, Nguyen22_Blueprint,Singh_2026_two-qubit_gates, stefanski2024readout}. We further fix $E_J/h = \qty{4.1}{\GHz}$, also representative of typical fluxonium qubits. With these parameters, the transition frequency at the symmetry point is \qty{400}{\MHz}, matching the $\cos(2\phi)$ qubit frequency in this work at the bias point $\Phi_{\rm ctrl}/\Phi_0 = 0.378$, as shown in Fig.~\ref{fig:Qubit_spec}(c).
Using the same noise sources and identical noise quality factors (see Tab.~\ref{tab:noise_params}), we simulate the corresponding lifetime of this qubit and find a significantly larger lifetime at the symmetry point of around \qty{130}{\us}. This is two orders of magnitude larger than in the $\cos(2\varphi)$ device studied in this work. However, we still find a lifetime near the symmetry point to be limited by $1/f$ flux noise, in agreement with experimental measurements \cite{ateshian2025, Bylander_2011}. The sensitivity of flux-based qubits to $1/f$ flux noise likely originates from local magnetic two-level system defects in the interface layers surrounding the SQUID loops and can be quantified by a corresponding noise strength $A_{1/f}$ \cite{Braum_ller_2020}. The noise amplitude $A_{1/f} = 1.5 \times 10^{-5}\,\Phi_0$ in our experiment is consistent with values typically observed for intrinsic flux noise \cite{ateshian2025, Yan_2016, Bylander_2011, Braum_ller_2020}. 

To understand the difference in flux noise sensitivity between the fluxonium qubit and the device in this work, we note that the coupling operator for $1/f$ flux noise scales with the inductive energy $E_L$ (in the few \qty{100}{\MHz} range), as follows from Eq.~\eqref{eq:Fluxonium_ham}. This reduced energy scale, compared with around \qty{40}{\GHz} for the $\cos(2\varphi)$ qubit, leads to a lower sensitivity to flux noise and correspondingly longer lifetimes. This lower sensitivity to $1/f$ flux noise similarly applies to recently demonstrated $\cos (2\varphi)$ qubits based on the kinetic interference co-tunneling element) (KITE)~\cite{roverch2026experimentalcos2phi}.

\section{Conclusion}

In summary, we fabricated a device based on Fourier engineering of the energy-phase relation to realize an interference-based protected qubit. The measured spectroscopy shows excellent agreement with the simplified theoretical model and reveals the appearance of higher harmonics in the qubit potential energy.

Time-domain measurements were performed to track the evolution of qubit lifetime as the device is tuned toward the protected regime. We find that the lifetime is generally limited by $1/f$ flux noise.
Although the device benefits from a relatively simple fabrication process, its coherence properties appear to be intrinsically limited by $1/f$ flux noise due to the high sensitivity associated with first-harmonic cancellation near the sweet spot and the associated noise scaling as the effective energy of the system $E_J$. At the same time, for comparable qubit frequencies at the symmetry point and assuming the same noise amplitude, the fluxonium qubit is predicted to exhibit a lifetime that is two orders of magnitude larger. Our results are consistent with the independent work presented in Ref.~\cite{rhombus2026}, where an asymmetric rhombus circuit is studied. On the other hand, the mechanism responsible for the observed upper limit of lifetime away from the symmetry point, which differs from the simulated behavior and remains consistent across different datasets, is not yet fully understood and is to be explored in future work.

\section*{Data and Code availability}
All numerical and experimental data are available through \cite{Data_repository} and the code used for the numerical simulations and data processing is available through \cite{GitHub_repository}.

\section*{Acknowledgments}
We acknowledge useful discussions with K.~Shagalov, D.~Feldstein-Bofill, S.~Kr{\o}jer, M.~Kjaergaard, P.A.~Sanchez and A.~Gyenis. This work was financially supported by the Dutch Research Council (NWO) and Holland High Tech (TKI) project 00PPS334.


\appendix
\renewcommand{\thefigure}{S\arabic{figure}}
\renewcommand{\theHfigure}{S\arabic{figure}}
\setcounter{figure}{0}

\section*{Supplementary Information}

\section{Fabrication}\label{sup:Fabrication}

Fabrication process used for the $\cos(2\varphi)$ circuit is similar to the fluxonium process described in Ref.~\cite{Singh_2026_two-qubit_gates}. The device is fabricated on a high-resistivity silicon substrate. Prior to metal deposition, the wafer is cleaned using nitric acid (HNO$_3$) followed by a 40\% hydrofluoric acid (HF) dip to remove the native oxide. Immediately after cleaning, the wafer is transferred to the deposition chamber to minimize surface reoxidation and contamination. 
The base superconducting film consists of a \qty{15}{\nm} Nb seed layer and a \qty{200}{\nm} Ta layer. Following base layer deposition, electron-beam lithography is used to define the circuit features, followed by reactive-ion etching (RIE) in an SF$_6$/O$_2$ plasma. The etch is performed in two steps: a high-power etch to ensure a straight etch profile, followed by a lower-power etch to ensure a smooth silicon surface.

Before Josephson junction fabrication, the sample undergoes a second cleaning step using nitric acid (HNO$_3$), followed by a buffered oxide etch (BOE 7:1) to remove surface oxides and contaminants. A trilayer resist stack of MMA/PMMA/PMMA is used for Josephson junction patterning, defining the required undercut profile. 
The Josephson junctions are fabricated using double-angle evaporation in an electron-beam evaporation system. The process consisted of two aluminum depositions separated by an in-situ oxidation step, forming the AlO$_x$ tunnel barrier. A 90$^\circ$ rotation between depositions define the junction geometry. The junctions were finalized using a standard lift-off process. To ensure good contact to the base-layer, we employ an additional aluminum patch. For the patch fabrication, a single-layer resist is used, with patch features defined by electron-beam lithography. Aluminum films are then deposited in the same electron-beam evaporation system. Prior to deposition, an ion-milling step is performed to remove native oxide and resist residues, ensuring good galvanic contact between the junction strip and the underlying metal layer.

The fabrication process described above determines the geometry and properties of the Josephson junctions, which may lead to variations in their effective Josephson energies. The discrepancy between $E_{J1,2}$ and $E_{J3}$ arises from differences in the junction geometries. The first two junctions are designed with identical dimensions (\qty{0.487}{\um} $\times$ \qty{0.487}{\um}) while the third junction is designed with the dimensions of \qty{0.368}{\um} $\times$ \qty{0.646}{\um}.The larger width of the third junction can lead to deposition on the sidewalls of the undercut, shadowing the junction and possibly increasing its effective width.

\section{Measurement setup}\label{sup:Measurement setup}

\begin{figure}[t]
    \centering
    \includegraphics[width=0.75\linewidth]{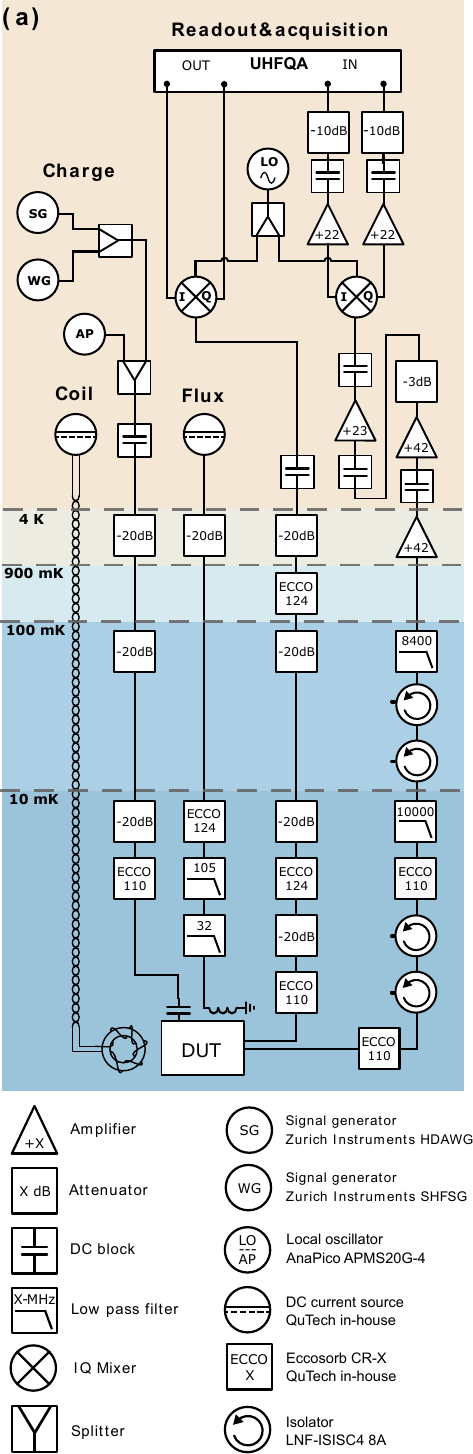}
    \caption{ Wiring diagram of experimental setup, see additional description in the text.}
    \label{fig_sup:setup}
\end{figure}

The experimental setup used for device characterization is similar to the one described in Ref.~\cite{Zwanenburg_2025, Singh_2026_two-qubit_gates}. Figure~\ref{fig_sup:setup} shows the full experimental wiring. All experiments are performed in a Bluefors LD400 dilution refrigerator at a base temperature of \qty{10}{\milli\kelvin}.
The global flux bias is provided by a superconducting coil mounted on the back of the sample enclosure. The sample is further protected from thermal and electromagnetic radiation by a copper can and two mu-metal shields. An in-house-built DC-current source is used to bias the flux line and the superconducting coil. Qubit drive pulses are generated using a Zurich Instruments HDAWG and Zurich Instruments SHFSG, combined with a microwave signal from an AnaPico APMS20G-4. The readout pulses are generated and analyzed by the Zurich Instruments UHFQA. The readout signals are upconverted using a Zurich Instruments HDIQ mixer with a local oscillator signal from the AnaPico APMS20G-4. All signals pass through a series of attenuators, filters, and in-house-made Eccosorb infrared filters. The output signal is amplified by a chain consisting of cryogenic dual-junction isolators, a cryogenic HEMT (LNF-LNC4\_8C), a room-temperature HEMT (LNF-LNR4\_8ART), and a 23~dB amplifier (Mini-Circuits ZRON-8G+), before demodulation.

\section{Cross-talk calibration}\label{Sup: Cross-talk calibration}

\begin{figure}[t]\label{fig_sup:Cross-talk calibration}
    \centering
    \includegraphics[width=0.5\textwidth]{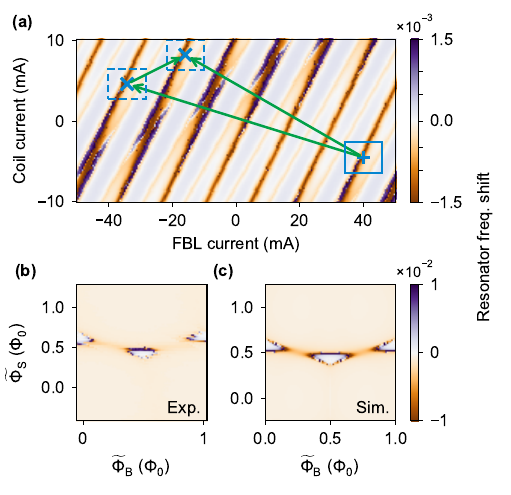}
    \caption{
    (a) Measurement results without cross-talk compensation, obtained by sweeping the currents applied to the coil and FBL. The solid rectangle shows the auto-correlation kernel used for identifying periodic features in the data, and the dashed rectangles are the detected matches. The extracted lattice vectors (green arrows) can be used to define a unit cell over which the heatmap can be periodically tiled.
    (b) Measured resonator spectroscopy after applying cross-talk correction, showing good agreement with the simulated results in (c).
    }
    \label{fig:cross_talk_calibration}
\end{figure}

The qubit is controlled by two magnetic fields: one generated by an on-chip flux-bias line (FBL) and the other by a coil mounted on the back of the sample holder. Both sources generate magnetic flux that threads the two loops simultaneously. To accurately compare measurements with the simulated Hamiltonian, flux cross-talk must be calibrated.
We first perform resonator spectroscopy as a function of both currents applied to the FBL and the coil, see Fig.~\ref{fig_sup:Cross-talk calibration}. We measure the resonator frequency shift, which provides a fast and accurate probe of the system properties. In Fig.~\ref{fig:cross_talk_calibration}(a), we observe a squeezed pattern that repeats with high periodicity, reflecting the combined tuning of the qubit by both strongly correlated flux sources.

To transform from the current basis to calibrated flux variables ($\Phi_{\rm bias}, \Phi_{\rm ctrl}$), that is, to calibrate out the cross-talk, we identify repeating features in the resonator frequency heatmap, as indicated by the squares in Fig.~\ref{fig_sup:Cross-talk calibration}(a). Using the unit cell defined by the basis vectors shown in the figure, we reconstruct the full periodic structure by tiling. We then compensate for the cross-talk using the following coordinate transformation matrix:
\begin{equation}
    \begin{matrix}\label{tab:rotation_matrix}
        & \text{FBL (mA)}& \text{Coil (mA)}\\
        \Phi_{\rm bias}/\Phi_0 & 0.0993\phantom{0} & 0.0307\phantom{0} \\
        \Phi_{\rm ctrl}/\Phi_0 & 0.14225 &  0.03525\rlap{\,.}
        \end{matrix}
\end{equation}
As seen in Fig.~\ref{fig:cross_talk_calibration}(b), the measured resonator spectrum after compensating using the characterized matrix is in good agreement with the simulated results. In this step, the linear compensation transforms the system to the intermediate basis $\widetilde{\Phi}_{\rm B} = {\Phi}_{\rm B} - \Phi_{\rm S}/2$, where only the $\Phi_{\rm S}/2$ contribution is compensated, rather than the full offset $\delta$. An additional transformation to the basis $\Phi_{\rm bias}$, where the sweet spot remains centered around $0.5\,\Phi_0$, is applied after spectroscopy and time-domain measurements.


\section{Calibration and measurement sequence}

To characterize the dependence of the lifetime on external flux $\Phi_{\mathrm{\rm bias}}$, measurements are performed at multiple flux points around $0.5\,\Phi_0$. As the system changes its properties, such as qubit frequency and coherence time, different measurement parameters must be updated for each flux point. To automate the measurement procedure, initial parameter estimates are obtained from a set of calibration measurements around the symmetry point. Subsequently, for each $\Phi_{\mathrm{\rm bias}}$ value, we execute a full measurement sequence, following the qubit frequency trace obtained from the spectroscopy measurements.

The measurement sequence consists of (i) resonator spectroscopy, (ii) continuous-wave two-tone spectroscopy, (iii) pulsed two-tone spectroscopy, (iv) Rabi oscillation measurements, and (v) time-domain measurements including $T_1$. At each flux point, the relevant parameters are extracted by fitting the data to the corresponding models and are used to calibrate subsequent measurements in the sequence. For the $T_1$ measurements, $2^{14}$ averages are acquired. Due to the simple measurement loop, not all parameters are perfectly optimized at every flux point, resulting in imperfections in the dataset. For further analysis, we therefore keep only $T_1$ values with fitted uncertainties below \qty{1}{\us}.

\section{Noise sources derivation}\label{sup: Noise}

To model the relaxation time of the qubit states through energy exchange with the environment, we use Fermi’s golden rule. Thus, the transition rate for each noise source is calculated using
\begin{equation}
    \Gamma_{1,\lambda}= \frac{1}{\hbar^2}|\bra{0}D_\lambda\ket{1}|^2S_\lambda (\omega_{10}),   
\end{equation}
where $D_\lambda$ is the operator associated with the noise source $\lambda$ and $S_\lambda (\omega_{10})$ is the noise spectral density at the qubit frequency.

\subsubsection{Dielectric loss} 

Dielectric noise comes from dissipation in lossy capacitive elements and can be modeled with the noise operator and an effective noise spectral density 
\begin{equation} 
    \begin{gathered}
    \hat{D}_{\rm diel} = 2e\hat{n}\,,\\
    S_{\rm diel}(\omega) = \frac{\hbar}{C_J Q_\mathrm{cap}(\omega)}\coth\left(\frac{\hbar|\omega|}{2k_BT}\right)\,,
    \end{gathered}    
\end{equation}
where $C_J$ is the relevant capacitance and $Q_{\mathrm{cap}}$ is the frequency-dependent capacitive quality factor. We assume a power-law dependence of the quality factor on frequency, referenced to $\omega_{\mathrm{ref}}/2\pi = 6\,\mathrm{GHz}$ \cite{Wang_2015, Nguyen2019_Fluxonium}
\begin{equation}
 Q_{\rm cap}(\omega) = A_{\rm cap}\left(\frac{2\pi \times \qty{6}{\GHz}}{|\omega|}\right)^\alpha.
\end{equation}
The parameters $A_{\mathrm{cap}}$ and $\alpha$ are treated as free fit parameters and are found to be $A_{\mathrm{cap}} = 1\times10^5$ and $\alpha = 0.7$. The extracted capacitive quality factor is around one order of magnitude lower than values in state-of-the-art experiments~\cite{Wang_2015}. This reduction could arise from fabrication-induced degradation of the substrate–metal and metal–air interfaces, for example, due to air exposure between processing steps or ion milling, as well as insufficient removal of surface contaminants.

\begin{figure}[t]
    \centering
    \includegraphics[width=0.99\linewidth]{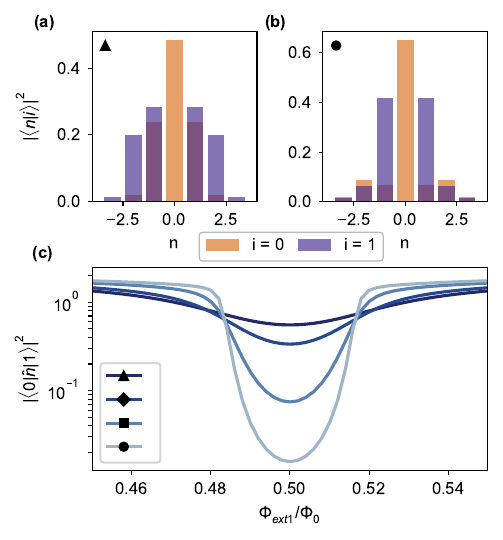}
    \caption{ Eigenstates and $0$--$1$ charge matrix element of the measured device. Panels 
    (a) and (b) show the ground and first excited state wavefunctions in the charge basis, for $\Phi_{\rm ctrl} = 0.406\,\Phi_0$ and $\Phi_{\rm ctrl} = 0.367\,\Phi_0$, respectively. As the system approaches the protected regime, the ground and first excited states develop an alternating charge-parity structure, with opposing parities. This reduces their wavefunction overlap, leading to a suppression of charge noise-induced transitions between the two states.
    (c) $0$--$1$ Charge matrix element for the four measured values of $\Phi_{\rm ctrl}$, demonstrating charge noise suppression, i.e., a reduction of the transition between states with different charge parity. 
    }
    \label{fig_sup:Charge_basis}
\end{figure}

The predicted improvement in coherence for the $\cos(2\varphi)$ protected qubit originates from the exponential suppression of charge matrix elements. As shown in Fig.~\ref{fig_sup:Charge_basis}, the overlap between the ground and first excited states decreases rapidly, reflecting their separation into states with different parity. This reduction in wavefunction overlap leads to a strong suppression of the charge matrix element $\bra{0}\hat{n}\ket{1}$, which reduces the coupling to charge noise dissipation channels. 

\subsubsection{Inductive noise} 

Inductive noise originates from dissipation in inductive elements. Although no explicit inductor is present in our system, we include this noise source to account for effective inductive loss arising from higher harmonics of internal modes~\cite{shagalov2025}. Inductive noise is described by the following noise operator and corresponding spectral density
\begin{equation} 
    \begin{gathered}
    \hat{D}_{\rm ind} =  \frac{\Phi_0}{2\pi}\hat{\varphi}\,,\\
      S_{\rm ind}(\omega) = \frac{\hbar}{L_{J} Q_{\rm ind}(\omega)}\coth\left(\frac{\hbar|\omega|}{2k_BT}\right)\,,
    \end{gathered}
\end{equation}
where $L_J$ is the effective inductance and $Q_{\mathrm{ind}}(\omega)$ is the frequency-dependent inductive quality factor.
To capture the frequency dependence of the inductive loss, we adopt the model of Ref.~\cite{Smith2020, Nguyen2019_Fluxonium}:
\begin{equation}
Q_{\rm ind}(\omega) =  A_{\rm ind} \frac{ K_{0} \left( \frac{h \times 0.5 {\rm GHz}}{2 k_B T} \right)
\sinh \left( \frac{h \times 0.5 {\rm GHz} }{2 k_B T} \right)}{K_{0} \left( \frac{\hbar |\omega|}{2 k_B T} \right)\
\sinh \left( \frac{\hbar |\omega| }{2 k_B T} \right)}\,,
\end{equation}
where $K_0$ is the modified Bessel function of the second kind. 
For superinductances consisting of a Josephson junction array, the quality factors have been estimated to be lower-bounded by $A_{\rm ind} = 5\times10^8$ at the frequency $\omega = 2\pi0.5$ GHz \cite{IoanPop}. In our calculations, we adopt this value and fix $A_{\rm ind} = 5 \times 10^8$.

\subsubsection{Ohmic flux-bias line noise} 

Ohmic flux noise arises from current fluctuations in the flux-bias line, which couple to the tunable qubit through the external magnetic flux. The corresponding noise operator and spectral density are modeled as
\begin{equation}
    \begin{gathered}
\hat{D}_{\rm FBL} = \frac{\partial \hat{H}}{\partial \Phi_x}\,,\\
    S_{\rm FBL}(\omega) = \frac{M^{2} \omega \hbar}{R} \coth\left(\frac{\hbar|\omega|}{2k_BT}\right)\,,
    \end{gathered}
\end{equation}
where $M$ is the mutual inductance between the qubit and the flux-bias line, and $R$ is the effective impedance of the bias line \cite{Transmon}. Since we have two loops in the system, both of them will contribute to the noise in the system through the operators
\begin{equation}
    \hat{D}_{\mathrm{\rm bias}} = \frac{\partial  \hat{H}}{\partial \Phi_{\rm bias}} = \frac{\partial \left( U_{R} + U_{R}^{\mathrm{int}}\right)}{\partial \Phi_{\rm bias}} 
\end{equation}
and
\begin{equation}
    \hat{D}_{\mathrm{\rm ctrl}} = \frac{\partial \hat{H}}{\partial \Phi_{\rm ctrl}} =  \frac{\partial \left( U_{R} + U_{R}^{\mathrm{int}}\right)}{\partial \Phi_{\rm ctrl}},
\end{equation}
where $U_R$ is the potential energy of the qubit circuit.
In our device, two flux degrees of freedom are present, corresponding to $\Phi_{\mathrm{\rm bias}}$ and $\Phi_{\mathrm{\rm ctrl}}$. Both contribute to relaxation through independent noise channels.
In the calculation, we include both contributions as the sum of the corresponding matrix elements
\begin{equation}
|\langle 0 | \hat{D}_{\mathrm{\rm bias}} | 1 \rangle|^2 + |\langle 0 | \hat{D}_{\mathrm{\rm ctrl}} | 1 \rangle|^2.
\end{equation}
This treatment assumes that the two noise sources are uncorrelated, which is a simplifying approximation for the system. Based on the loop geometry, the mutual inductance is estimated to be $M_{1,2} = 1800\,\Phi_0/A$.

\subsubsection{Quasiparticle-tunneling noise}

Quasiparticle tunneling generates current fluctuations through single-electron tunneling across the junction. These fluctuations couple to the qubit phase degree of freedom and are described by the noise operator and spectral density \cite{Catelani_2011, IoanPop, ateshian2025}
\begin{equation} 
    \begin{gathered}
    \hat{D}_{\rm qp}^F = (\Phi_0/\pi)\sin\left(\hat{\varphi}/2\right)\,,\\
      S_{\rm  qp}^F(\omega) = \hbar \omega\mathrm{Re}[Y_{\rm qp(\omega)}]\coth\left(\frac{\hbar|\omega|}{2k_BT}\right)\,,
    \end{gathered}
\end{equation}
The dissipative part of the quasiparticle admittance is given by
\begin{align}
  \mathrm{Re}[Y_{\rm qp(\omega)}] ={}& x_\mathrm{qp}\sqrt{\frac{2}{\pi}}\frac{8E_J}{R_K\Delta}\left(\frac{2\Delta}{\hbar\omega}\right)^{3/2} \times
  \\
  &\sqrt{\frac{\hbar\omega}{2k_BT}}
  K_0\left(\frac{\hbar|\omega|}{2k_BT} \right)\sinh\left(\frac{\hbar\omega}{2k_BT}\right)
  \nonumber
\end{align}
where $\Delta$ is the aluminum superconducting gap, $R_K = h/e^2$ is the resistance quantum, and $K_0$ is the modified Bessel function of the second kind. 
The quasiparticle density is taken to be $x_{\mathrm{qp}} = 7 \times 10^{-10}$ for both junctions. This value is two orders of magnitude lower than observed in previous experiments~\cite{IoanPop}, which may be attributed to improved shielding of the device. In this regime, the device is not expected to be limited by quasiparticle noise. 

In the main analysis of the fluxonium qubit, we evaluate quasiparticle tunneling across the small junction. We neglect quasiparticle contributions from the junction array. As discussed in Ref.~\cite{ateshian2025}, this contribution exhibits a flux dependence similar to that of inductive and flux noise near the half-flux bias point, making it difficult to distinguish experimentally.

\begin{figure}[t]
    \centering
    \includegraphics[width=0.5\textwidth]{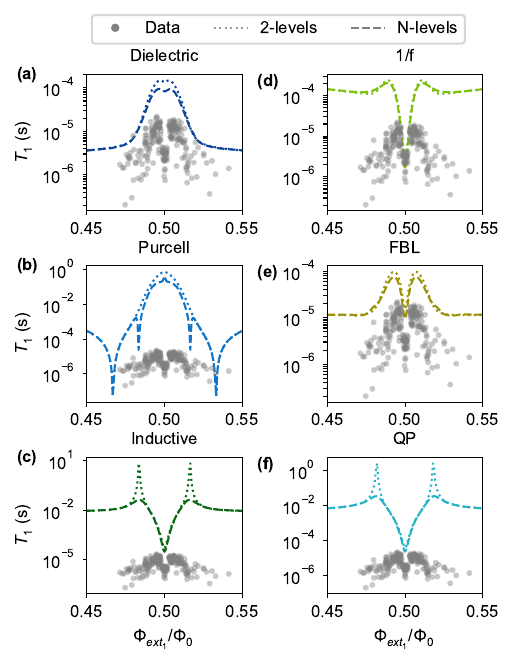}
    \caption{ Comparison of the two-level and $N$-level models with experimental data at a fixed $\Phi_{\rm ctrl} = 0.378\,\Phi_0$. Including higher energy levels does explain the observed limitation of $T_1$ near the sweet spot.
    }
    \label{fig:Nlevels}
\end{figure}

\subsubsection{Purcell decay to the resonator}

When the qubit is coupled to the resonator, the eigenstates become hybridized, acquiring a small photonic component. This leads to an additional decay channel, as the photonic component decays at the cavity loss rate $\kappa$, resulting in an effective qubit relaxation process \cite{Transmon, Blais_2004, Houck_2007}
In the far-detuned regime, where $|\Delta_0| \gg g_{01}$, the Purcell decay rate is given by
\begin{equation} 
    \begin{split}
  \Gamma^{01}_\kappa = \kappa \left(\frac{g_{01}}{\Delta_0}\right)^2
    \end{split}    
\end{equation}
where $\Delta_0$ is the detuning between the qubit $0 \rightarrow 1$ transition and the resonator frequency, defined as $\Delta_0 = \omega_{10} - \omega_\mathrm{res}$.

The cavity decay rate $\kappa$ is extracted from resonator measurements by fitting the resonance response to determine the total (loaded) quality factor $Q_{L}$, such that 
\begin{equation}
    \kappa = \omega_\mathrm{res} / Q_{L}.
\end{equation}
The qubit–resonator coupling strength is given by 
\begin{equation}
    g_{01} = g |\langle0 |\hat{n}|1\rangle|
\end{equation}
where $g$ is the geometrical coupling strength determined from electromagnetic simulations, and $\hat{n}$ is a charge operator. From resonator spectroscopy measurements (Fig.~\ref{fig:Resonator_spec}), we extract the bare resonator frequency $\omega_\mathrm{res}/2\pi = \qty{5.344}{\GHz}$, as well as the coupling strength $g/2\pi = \qty{25}{\MHz}$ from the resonator shift (Eq.~\eqref{eq:res_shift}).

\subsubsection{$1/f$ flux noise}

One of the dominant noise channels in superconducting qubits is low-frequency $1/f$ flux noise that couples to the phase degree of freedom. Its operator and spectral density are modeled as
\begin{equation} \label{eq:1/f_flux_noise}
    \begin{gathered}
\hat{D}_{\rm 1/f} = \frac{\partial \hat{H}}{\partial \Phi_x}\,,\\
    S_{\rm 1/f}(\omega) = \frac{2 \pi A_{1/f}^{2} }{|\omega|}\,,
    \end{gathered}
\end{equation}
where $A_{1/f}$ characterizes the amplitude of the flux noise~\cite{ateshian2025}. This noise is concentrated at low frequencies and primarily leads to slow fluctuations of the qubit transition frequency, resulting in dephasing. However, when the phase operator couples the $\ket{0}$ and $\ket{1}$ states and the qubit operates in the low-frequency regime where $1/f$ noise is strong, such fluctuations can contribute to energy relaxation \cite{Groszkowski_2018}.
In our device, we extract a flux noise amplitude of $A_{1/f} = 1.5 \times 10^{-5}\Phi_0$.

\subsection{$N$-Level Decoherence Model} \label{Higher level contribution}

To investigate the origin of the coherence limit in our qubit, we use a model that accounts for transitions involving higher energy levels, following approaches developed for fluxonium qubits~ \cite{ateshian2025, azar2026characterizationenergyrelaxation}. We employ a rate-matrix model to numerically evaluate population transfer between the first $N$ energy levels, with populations encoded in a time-dependent vector $\vec{p}(t)$. The time evolution is governed by $\partial_t \vec{p}(t) = \boldsymbol{B}\vec{p}(t)$ where $\boldsymbol{B}$ is an $N \times N$ rate matrix, given as
\begin{equation}
\boldsymbol{B} = \begin{pmatrix}
        -\sum_i \Gamma_{0\rightarrow i} & \Gamma_{1 \rightarrow 0} & \cdots & \Gamma_{N \rightarrow 0} \\
        \Gamma_{0\rightarrow 1} & -\sum_i \Gamma_{1\rightarrow i} & \cdots & \vdots\\
        \vdots & \vdots & \ddots & \Gamma_{N\rightarrow N-1}\\
        \Gamma_{0\rightarrow N} & \Gamma_{1\rightarrow N} & \cdots & -\sum_i \Gamma_{N\rightarrow i}
        \end{pmatrix}
\end{equation}
where the transition rates $\Gamma_{i\rightarrow j}$ are calculated using Fermi’s golden rule for the corresponding noise source. In this model, we assume the system is in thermal equilibrium with the environment, such that the upwards transition is $\Gamma_{i\rightarrow j} = \Gamma_{j\rightarrow i}{\rm exp}\left(\hbar(\omega_i-\omega_j)/k_BT\right)$. The matrix
$B$ can be diagonalized as $\boldsymbol{B} = \boldsymbol{VSV}^{-1}$, where $\boldsymbol{S}$ is a diagonal matrix with the eigenvalues on the diagonal and $\boldsymbol{V}$ is the matrix built from the eigenvectors as columns of matrix $\boldsymbol{B}$. The solution to the population dynamics is then given by $\vec{p}(t) = e^{\boldsymbol{B}t}\vec{p}(0) = \boldsymbol{V}e^{\boldsymbol{S}t}\boldsymbol{V}^{-1}\vec{p}(0)$. 
In all simulations, we assume an effective qubit temperature of 40~mK and include the lowest $N$ energy levels, verifying that higher levels do not significantly affect the results.

As shown in Fig.~\ref{fig:Nlevels}, including higher levels primarily affects dielectric and Purcell loss, leading to a reduction in the qubit lifetime. However, these mechanisms alone do not fully account for the measured $T_1$, indicating that additional relaxation channels may be present.
\newpage
\bibliography{reference}

\end{document}